\documentclass[a4paper,aps,10pt,prd,preprintnumbers,showpacs,twocolumn,superscriptaddress,nofootinbib,amsmath,amssymb]{revtex4-1}
\usepackage{graphicx}
\DeclareMathAlphabet{\pazocal}{OMS}{zplm}{m}{n}

\begin{document}
\title{Hawking radiation of non-Schwarzschild black holes in higher derivative gravity: a crucial role of grey-body factors}
\author{R. A. Konoplya}
\email{roman.konoplya@gmail.com}
\affiliation{Institute of Physics and Research Centre of Theoretical Physics and Astrophysics, Faculty of Philosophy and Science, Silesian University in Opava, CZ-746 01 Opava, Czech Republic}
\affiliation{Peoples Friendship University of Russia (RUDN University), 6 Miklukho-Maklaya Street, Moscow 117198, Russian Federation}
\author{A. F. Zinhailo}\email{F170631@fpf.slu.cz}
\affiliation{Institute of Physics and Research Centre of Theoretical Physics and Astrophysics, Faculty of Philosophy and Science, Silesian University in Opava, CZ-746 01 Opava, Czech Republic}
\begin{abstract}
The higher derivative gravity includes corrections of the second order in curvature and allows for both Schwarzschild and non-Schwarzschild asymptotically flat black-hole solutions. Here we find the grey-body factors and energy emission rates for Hawking radiation of test Dirac and electromagnetic fields in the vicinity of such a non-Schwarzschild black hole. The temperature and mass of the black hole monotonically decrease from its Schwarzschild value to zero when the coupling constant is increased up to its extremal value. Nevertheless, for small and moderate values of the coupling constant the Hawking radiation is enhanced, and only in the regime of large coupling it is suppressed, as one could expect. The reason for such counter-intuitive behavior is the important role of the grey-body factors: for small and moderate couplings the temperature falls relatively slowly, while the effective potentials for black holes of the same mass become considerably lower, allowing for much higher transmission rates. We have also estimated the lifetime of such black holes and shown that the range of black-hole masses at which ultra-relativistic emission of massive electrons and positrons starts is shifted towards smaller black-hole masses when the coupling constant is large.
\end{abstract}
\maketitle

\section{Introduction}

The new era in observations of black holes in the gravitational \cite{Abbott:2016blz} and electromagnetic \cite{Akiyama:2019cqa,Goddi:2016jrs} spectra gave an opportunity to test the regime of strong gravity. Even though the General Relativity is most favorable and celebrated candidate, there are a number of fundamental questions which could be answered by modifying the Einstein gravity at high energies. These are construction of quantum gravity, the singularity problem, dark energy and dark matter problems, the hierarchy problem. After all, the current uncertainty in determination the angular momenta and masses of  the observed black holes leaves rather large parametric freedom for alternative theories of gravity \cite{Konoplya:2016pmh,Berti:2018vdi}.

An essential problem when constructing quantum theory of gravity is the non-renormalizability of a theory. A general approach to solution of the non-renormalizability of General Relativity is connected with adding higher order terms to the theory \cite{tHooft}. One of such effective theories of gravity at low energies is the Einstein gravity with added quadratic in curvature term, which has the following form:

\begin{equation}
\label{HDGaction}
I = \int d^4x\sqrt{-g}\left(\gamma R -
\alpha C_{\mu\nu\rho\sigma}C^{\mu\nu\rho\sigma} + \beta R^{2}\right)\,,
\end{equation}
where $\alpha$, $\beta$ and $\gamma$ are coupling constants, $C_{\mu\nu\rho\sigma}$ is the Weyl tensor. For spherically symmetric and asymptotically flat solutions we can choose $\gamma =1$ and $\beta =0$ \cite{Lu:2015cqa,Lu:2015psa}, so that the only new coupling constant is $\alpha$. The condition $R=0$ is evidently satisfied in this case, so that the Schwarzschild solution is also the solution of the above theory (\ref{HDGaction}). In addition, there is an asymptotically flat non-Schwarzschild solution \cite{Lu:2015cqa}, which describes static spherically symmetric black hole. Thus, the theory (\ref{HDGaction}) allows for the two realizations of black holes. Following  \cite{Kokkotas:2017zwt} we will use the dimensionless parameter, which parameterizes the non-Schwarzschild solutions up to the rescaling
\begin{equation}\label{111}
p=\frac{r_0}{\sqrt{2\alpha}},
\end{equation}
where $r_{0}$ is the radius of the black-hole event horizon.

The numerical solution for the non-Schwarzschild black hole obtained in \cite{Lu:2015cqa,Lu:2015psa} was later represented in analytical form in \cite{Kokkotas:2017zwt} by using the generic parametrization of spherically symmetric black-hole spacetimes \cite{Rezzolla:2014mua}. The spherical parametrization \cite{Rezzolla:2014mua} was extended to the axially-symmetric case in \cite{Konoplya:2016jvv}. The non-Schwarzschild solution has an additional parameter $p$ related to the coupling constant $\alpha$ according to (\ref{111}). This solution has one interesting property: when $p$ increases from its minimal value $p = p_{min}$, the mass of the black hole is monotonically decreasing until zero in the extremal case $p=p_{max}$.
The radius of the extremal black hole does not vanish.

Various aspects of black-hole physics in the higher derivative theories have been recently considered in \cite{Bonanno:2019rsq,Sultana:2019lhf,Salvio:2019ewf,Cano:2019ore,Svarc:2018coe,Lin:2016kip,Podolsky:2018pfe}.
The quasirnomal modes for test scalar and electromagnetic field were calculated with sufficient accuracy in \cite{Zinhailo:2018ska}. Nevertheless, the Hawking radiation and scattering problem for such black holes have not been considered so far.
At the same time, the Hawking radiation is known to be more sensitive characteristic then quasinormal modes when the higher order corrections are taken into consideration \cite{Konoplya:2019hml}. For higher dimensional black holes even the relatively small quadratic (Gauss-Bonnet) correction leads to strong suppression of Hawking radiation by several orders \cite{Konoplya:2010vz,Rizzo:2006uz}.
Therefore, taking into account unusual properties of the non-Schwarzschild solution in the higher derivative gravity on one side and trying to understand how general is the strong suppression of Hawking radiation under the higher curvature corrections on the other, here we would like to evaluate the intensity of Hawking evaporation for the above black holes.

As a result, we have found a number of new features of Hawking radiation in the vicinity the non-Schwarzschild black holes, main of which is the presence of two different regimes of Hawking radiation. Regardless the monotonic fall off of the black-hole temperature, the intensity of Hawking radiation is increased up to some critical value of the coupling constant $p$. We show that this happens, because the grey-body factors are more important than the temperature, when the coupling $p$ is far from its maximal value.

Our paper is organized as follows. Sec. II is devoted to equations of motion for test Maxwell and Dirac fields in the background of the non-Schwarzschild black hole in the Einstein-Weyl gravity. Sec. III tests the applicability of the WKB formula at higher orders to finding the grey-body factors of black holes by comparison the numerical and WKB data for the Schwarzschild case. Sec. IV is devoted to calculations of the grey-body factors, total energy emission rates for each field and estimation of the black-hole lifetime. Finally, in the Conclusions we summarize the obtained results and mention a number of open problems.

 \section{The metric and waves equations for test electromagnetic and Dirac fields}

\begin{figure}
\resizebox{\linewidth}{!}{\includegraphics*{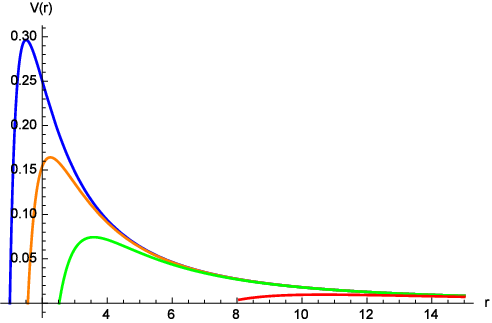}}
\caption{The effective potential $V(r)$ for the electromagnetic field ($s=1$), for $\ell=1$ in
the units $M= 1/2$, blue line corresponds to $p=0.876$, it is Schwarzschild limit, orange line corresponds to $p=0.975$, green line corresponds to $p=1.041$, red line corresponds the maximal value  $p=1.14$}\label{fig:1A}
\end{figure}
\begin{figure}
\resizebox{\linewidth}{!}{\includegraphics*{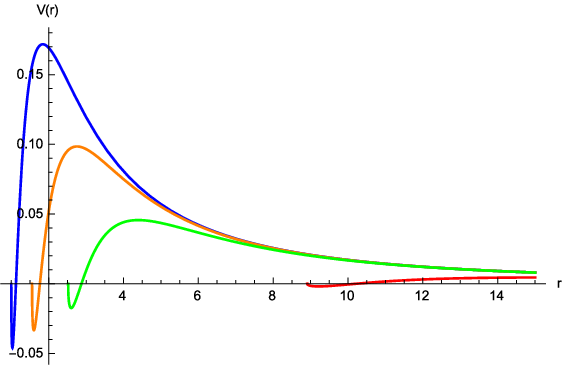}}
\caption{The effective potential $V_{-1/2}(r)$ for the Dirac field ($s=1/2$), for $\ell=1$ in
the units $M= 1/2$, blue line corresponds to $p=0.876$, it is Schwarzschild limit, orange line corresponds to $p=0.975$, green line corresponds to $p=1.041$, red line corresponds the maximal value $p=1.14$}\label{fig:1B}
\end{figure}

A static spherically symmetric metric has the general form:
\begin{eqnarray}\label{metric}
ds^2 &=& -e^{\mu}dt^2+e^{\nu}{dr^2}+r^2 (\sin^2 \theta d\phi^2+d\theta^2),
\end{eqnarray}
where the explicit expression for the metric coefficients were obtained numerically in \cite{Lu:2015cqa,Lu:2015psa} and approximated in an analytical form in \cite{Kokkotas:2017zwt}. Here we will use the analytical form suggested in  \cite{Kokkotas:2017zwt}, for which  the maximal relative error is about $0.06 \%$. The analytical expressions for the metric coefficients are given in the appendix.

An electromagnetic field obeys the general covariant Maxwell equations:
\begin{equation}\label{genelm}
\dfrac{1}{\sqrt{-g}} \partial_\mu(F_{\rho\sigma}g^{\rho\nu}g^{\sigma\mu}\sqrt{-g})=0.
\end{equation}
Here $A_{\mu}$ is a vector potential and $F_{\rho\sigma}=\partial_\rho A^\sigma - \partial_\sigma A^\rho$.
The general covariant Dirac equation has the form \cite{Brill:1957fx},
\begin{equation}
\gamma^{\alpha} \left( \frac{\partial}{\partial x^{\alpha}} - \Gamma_{\alpha} \right) \Psi  =0,
\end{equation}
where $\gamma^{\alpha}$ are noncommutative gamma matrices and $\Gamma_{\alpha}$ are spin connections in the tetrad formalism. After separation of angular variables, the wave equation can be represented in the following general master form (see, for instance \cite{Konoplya:2006rv,Zinhailo:2018ska,Brill:1957fx} and references therein):
\begin{equation}  \label{klein-Gordon}
\dfrac{d^2 \Psi_{s}}{dr_*^2}+(\omega^2-V_{s}(r))\Psi_{s}=0,
\end{equation}
where the relation
$$dr_*=\sqrt{e^{\nu-\mu}}dr$$
defines the ``tortoise coordinate'' $r_*$.
For all $p$ the Schwarzschild metric is the exact solution of the Einstein-Weyl equations as well, but only at some minimal nonzero $p_{min}$, in addition to the Schwarzschild solution, there appears the non-Schwarzschild branch which describes the asymptotically flat black hole, whose mass is decreasing, when $p$ grows. The approximate maximal and minimal values of $p$ are:
\begin{equation}
p_{min} \approx 1054/1203 \approx 0.876, \quad p_{max} \approx 1.14.
\end{equation}

The deviation of the black-hole radius from the Schwarzschild value $r = 2 M$ is expressed as follows:
\begin{equation}
\frac{2M}{r_0}-1 \approx (1054 - 1203 p)\left(\frac{3}{1271} + \frac{p}{1529}\right).
\end{equation}

\begin{figure}
\resizebox{\linewidth}{!}{\includegraphics*{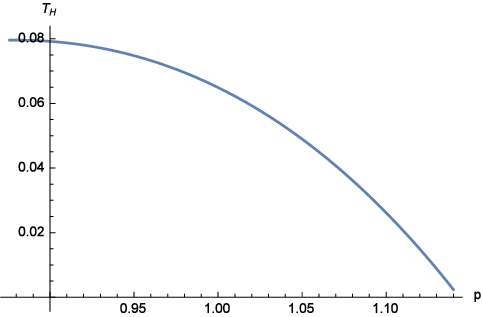}}
\caption{The Hawking temperature $T_H$ as a function of $p$ in the units $M=1/2$.}\label{fig:Temperature}
\end{figure}

The effective potentials of test Dirac ($s=\pm 1/2$) and electromagnetic ($s=1$) fields in the general background (\ref{metric}) can be written in the following forms:
%
\begin{equation}
V_{\pm 1/2} = \frac{k}{r}\left(\frac{e^{\mu}k}{r}\mp\frac{e^{\mu}\sqrt{e^{\nu}}}{r}\pm\sqrt{e^{\mu-\nu}}(\sqrt{e^{\mu}})'\right),
\end{equation}
\begin{equation}\label{scalarpotential}
V_{1}=\dfrac{\ell(\ell+1)e^{\mu}}{r^2}.
\end{equation}
Here $\ell = 1, 2...$ for the electromagnetic field and $k =\ell+1$ ($k =\ell$) for the ``plus'' (``minus'') potential of the Dirac field. In the both cases $k = 1, 2...$. The effective potentials for the electromagnetic field have the form of positive definite potential barriers (see fig. (\ref{fig:1A})), but the $V_{-1/2}$ potential of the Dirac field has a negative gap near the event horizon.

\section{Applicability of the 6th order WKB formula for evaluation of Hawking radiation}

We will consider the wave equation (\ref{klein-Gordon}) with the boundary conditions allowing for incoming waves from infinity. Owing to the symmetry of the scattering properties this is identical to the scattering of a wave coming from the horizon. The scattering boundary conditions for (\ref{klein-Gordon}) have the following form
\begin{equation}\label{BC}
\begin{array}{ccll}
    \Psi &=& e^{-i\omega r_*} + R e^{i\omega r_*},& r_* \rightarrow +\infty, \\
    \Psi &=& T e^{-i\omega r_*},& r_* \rightarrow -\infty, \\
\end{array}%
\end{equation}
where $R$ and $T$ are the reflection and transmission coefficients.
\begin{equation}\label{1}
\left|T\right|^2 + \left|R\right|^2 = 1.
\end{equation}
Once the reflection coefficient is calculated, we can find the transmission coefficient for each multipole number $\ell$
\begin{equation}
\left|{\pazocal
A}_{\ell}\right|^2=1-\left|R_{\ell}\right|^2=\left|T_{\ell}\right|^2.
\end{equation}
\begin{equation}\label{moderate-omega-wkb}
R = (1 + e^{- 2 i \pi K})^{-\frac{1}{2}},
\end{equation}
where $K$ can be determined from the following equation:
\begin{equation}
K - i \frac{(\omega^2 - V_{0})}{\sqrt{-2 V_{0}^{\prime \prime}}} - \sum_{i=2}^{i=6} \Lambda_{i}(K) =0.
\end{equation}
Here $V_0$ is the maximum of the effective potential, $V_{0}^{\prime \prime}$ is the second derivative of the
effective potential in its maximum with respect to the tortoise coordinate $r_{*}$, and $\Lambda_i$  are higher order WKB corrections which depend on up to $2i$th order derivatives of the effective potential at its maximum \cite{WKBorder} and $K$. This approach at the 6th WKB order was used for finding transmission/reflection coefficients of various black holes and wormholes in \cite{Konoplya:2019hml,Konoplya:2010kv,Volkel:2019ahb}, however no systematic consideration of the accuracy of this formula for evaluation of Hawking radiation was done. Therefore, here we will compare the energy emission rates obtained numerically in \cite{Page:1976df} and compare them with those given by the 6th order WKB formula.

\begin{table}
\begin{tabular}{|c|c|c|}
   \hline
  \hline
  $\ell$/$k$ & $d E/d t$ by Page & $d E/d t$ by WKB \\
  \hline
   \hline
\multicolumn{3}{|c|}{Neutrinos} \\
    \hline
     \hline
  $k=1$ & 0.000630 & 0.000639 \\
    \hline
  $k=2$ & 0.000024 & 0.000024 \\
    \hline
  $k=3$ & $4 \times 10^{-7}$ & $4.7 \times 10^{-7}$ \\
   \hline
  The~sum & 0.00065 & 0.00066 \\
  \hline
   \hline
\multicolumn{3}{|c|}{Photons} \\
    \hline
     \hline
  $\ell=1$ & $1.32 \cdot10^{-4}$ & $1.34 \cdot10^{-4}$ \\
  \hline
  $\ell=2$ & $2.80 \cdot10^{-6}$ & $2.67\cdot10^{-6}$ \\
    \hline
  $\ell=3$ & $4 \cdot10^{-8}$ & $4.03\cdot10^{-8}$ \\
    \hline
  The~sum & $1.35 \cdot10^{-4}$  & $1.37 \cdot10^{-4}$ \\
  \hline
   \hline
   \multicolumn{3}{|c|}{Gravitons} \\
    \hline
     \hline
  $\ell=2$ & $0.152 \cdot10^{-4}$ & $0.152 \cdot10^{-4}$ \\
  \hline
  $\ell=3$ & $1.6 \cdot10^{-7}$ & $1.43 \cdot10^{-7}$ \\
    \hline
  The~sum &  $0.154 \cdot10^{-4}$  & $0.154 \cdot10^{-4}$  \\
  \hline
   \hline
\end{tabular}
\caption{Energy emission for the Schwarzschild background ($M=1/2$) after integrating over all the quantum numbers and $\omega$ in Page's work \cite{Page:1976df} versus that obtained here with the help of the 6th order WKB formula.}
\label{Table1}
\end{table}

We will assume that the black hole is in the thermal equilibrium with its environment in the following sense: the temperature of the black hole does not change between emissions of two consequent particles. This implies that the system can be described by the canonical ensemble (see \cite{Kanti:2004nr} for a review). Therefore, the energy emission rate for Hawking radiation has the form \cite{Hawking:1974sw}:
\begin{align}\label{energy-emission-rate}
\frac{\text{d}E}{\text{d} t} = \sum_{\ell}^{} N_{\ell} \left| \pazocal{A}_l \right|^2 \frac{\omega}{\exp\left(\omega/T_\text{H}\right)\pm1} \frac{\text{d} \omega}{2 \pi},
\end{align}
were $T_H$ is the Hawking temperature, $A_l$ are the grey-body factors, and $N_l$ are the multiplicities, which only depend on the space-time dimension and $l$. The Hawking temperature for spherically symmetric black hole is
\begin{equation}
T_H = \frac{1}{4 \pi} \sqrt{-\frac{g^{\prime}_{tt}}{g^{\prime}_{rr}}}\bigg|_{r=r_{0}}.
\end{equation}
The multiplicity factors for the four dimensional spherically symmetrical black holes case consists from
the number of degenerated $m$-modes (which are $m = -\ell, -\ell+1, ....-1, 0, 1, ...\ell$, that is  $2 \ell +1$ modes) multiplied by the number of species of particles which depends also on the number of polarizations and helicities of particles. Therefore, we have
\begin{equation}
N_{\ell} = 4 (\ell+1) \qquad (Maxwell),
\end{equation}
\begin{equation}
N_{\ell} = 8 k \qquad (Dirac).
\end{equation}
The multiplicity factor for the Dirac field is fixed taking into account both the ``plus'' and ``minus'' potentials which are related by the Darboux transformations, what leads to the iso-spectral problem \cite{Zhidenko:2003wq,Konoplya:2012df} and the same grey-body factors for both chiralities. We will use here the ``minus'' potential, because the WKB results are more accurate for that case in the Schwarzschild limit.

\begin{figure}
\resizebox{\linewidth}{!}{\includegraphics*{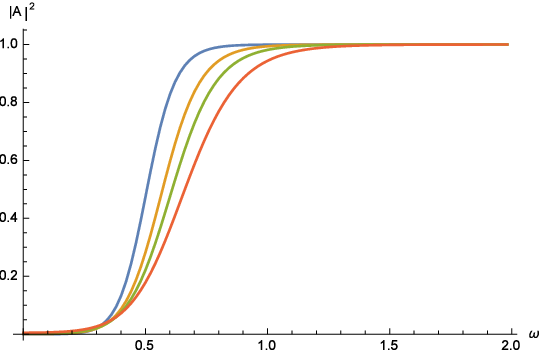}}
\caption{Grey-body factors as functions of $\omega$ for the electromagnetic field ($s=1$) for $\ell=1$, $r_{0}=1$. Blue line corresponds to $p=0.876$, it is Schwarzschild limit, orange line corresponds to $p=0.975$, green line corresponds to $p=1.041$, red line is $p=1.14$}\label{fig:1a}
\end{figure}

\begin{figure}
\resizebox{\linewidth}{!}{\includegraphics*{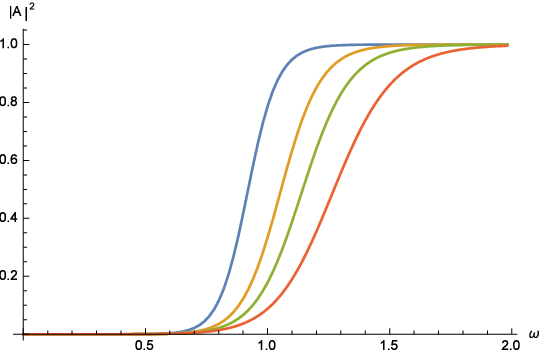}}
\caption{Grey-body factors as functions of $\omega$ for the electromagnetic field ($s=1$) for $\ell=2$, $r_{0}=1$. Blue line corresponds to $p=0.876$, it is Schwarzschild limit, orange line corresponds to $p=0.975$, green line corresponds to $p=1.041$, red line is  $p=1.14$}\label{fig:2a}
\end{figure}

\begin{figure}
\resizebox{\linewidth}{!}{\includegraphics*{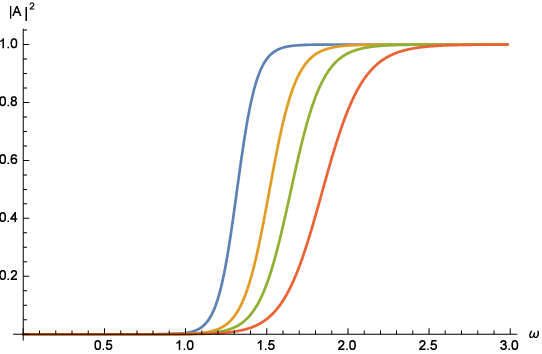}}
\caption{Grey-body factors as functions of $\omega$ for the electromagnetic field ($s=1$) for $\ell=3$, $r_{0}=1$. Blue line corresponds to $p=0.876$, it is Schwarzschild limit, orange line corresponds to $p=0.975$, green line corresponds to $p=1.041$, red line is  $p=1.14$}\label{fig:3a}
\end{figure}
From table (\ref{Table1}) one can see that the relative error related to the usage of the 6th order WKB formula is always less than $2\%$ which means very good approximation to us, as we will see that the total effect due to the parameter $p$ is much larger then the error.

The negative gap in the effective potential $V_{-1/2}$ of the Dirac field, which we observed, raises the other problem: is Dirac field classically stable or there are some bound states with negative energy? Although the depth of the negative gap is smaller  when the coupling $p$ is turned on, the potential barrier is lower, what could leave the possibility of instability. However, the positivity of the effective potential with the opposite chirality and the fact that both potentials are related by the Darboux transformations proves the stability of the Dirac field.

\section{Grey-body factors and energy emission rates}

\begin{figure}
\resizebox{\linewidth}{!}{\includegraphics*{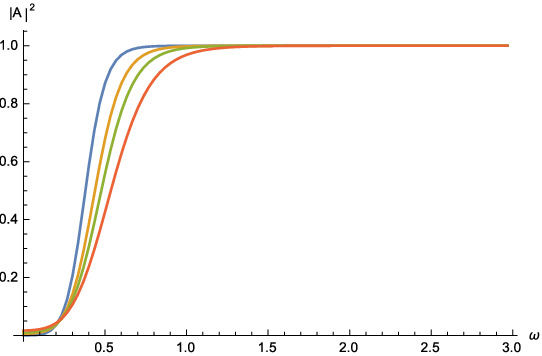}}
\caption{Grey-body factors as functions of $\omega$ for the Dirac field ($s=1/2$) for $k=1$, $r_{0}=1$. Blue line corresponds to $p=0.876$, it is Schwarzschild limit, orange line corresponds to $p=0.975$, green line corresponds to $p=1.041$, red line is $p=1.14$}\label{fig:1b}
\end{figure}

\begin{figure}
\resizebox{\linewidth}{!}{\includegraphics*{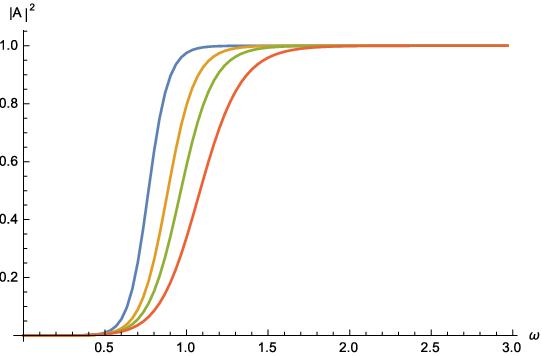}}
\caption{Grey-body factors as functions of $\omega$ for the Dirac field ($s=1/2$) for $k=2$, $r_{0}=1$. Blue line corresponds to $p=0.876$, it is Schwarzschild limit, orange line corresponds to $p=0.975$, green line corresponds to $p=1.041$, red line is  $p=1.14$}\label{fig:2b}
\end{figure}

\begin{figure}
\resizebox{\linewidth}{!}{\includegraphics*{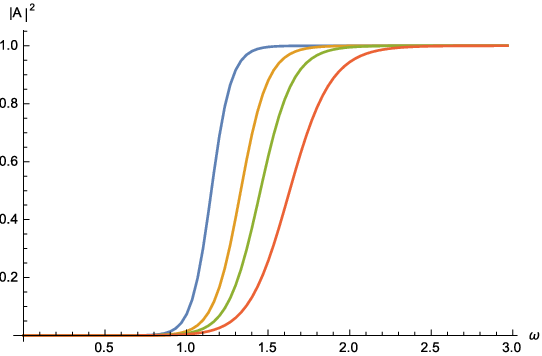}}
\caption{Grey-body factors as functions of $\omega$ for the Dirac field ($s=1/2$) for $k=3$, $r_{0}=1$. Blue line corresponds to $p=0.876$, it is Schwarzschild limit, orange line corresponds to $p=0.975$, green line corresponds to $p=1.041$, red line is $p =1.14$}\label{fig:3b}
\end{figure}

The grey-body factors computed with the help of the higher order WKB formula cannot be sufficiently accurate for very small $\omega$ as the turning points are well separated in this case and the Taylor expansion in the middle region becomes bad approximation. Fortunately, comparison with the accurate results for the Schwarzschild solution, shows that very small frequencies do not contribute in the energy emission rate seemingly.
The grey-body factors for photons are shown on figs. (\ref{fig:1a},\ref{fig:2a},\ref{fig:3a}) and for the Dirac field on figs. (\ref{fig:1b},\ref{fig:2b},\ref{fig:3b}). There one can see that even in the units of the black hole radius the grey-body factors decrease when the the coupling parameter $p$ is increased for the both fields and all values of the multipole number $\ell$.


\begin{figure}
\resizebox{\linewidth}{!}{\includegraphics*{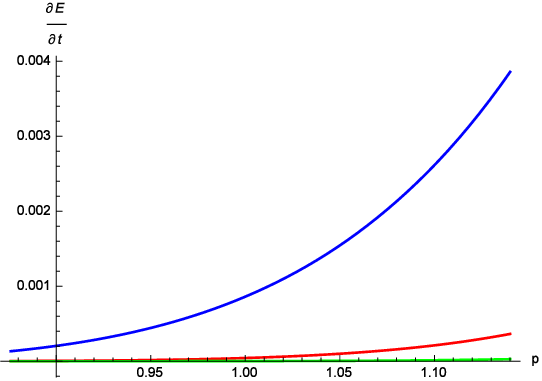}}
\caption{Energy emission of photons for the three multipoles $\ell =1, 2, 3$ (from top to the bottom) in the units  $r_{0}=1$.}\label{fig:4a}
\end{figure}

\begin{figure}
\resizebox{\linewidth}{!}{\includegraphics*{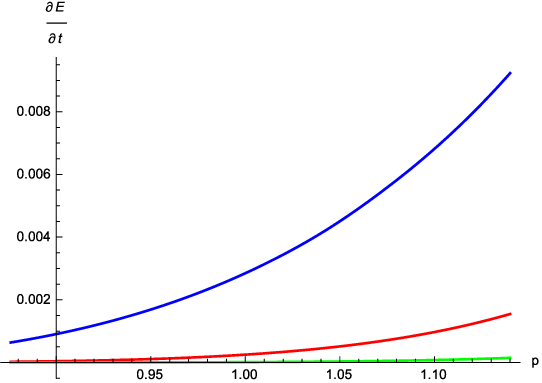}}
\caption{Energy emission of neutrinos for the three multipoles $\ell =1, 2, 3$ (from top to the bottom) in the units  $r_{0}=1$.}\label{fig:4b}
\end{figure}

\begin{figure}
\resizebox{\linewidth}{!}{\includegraphics*{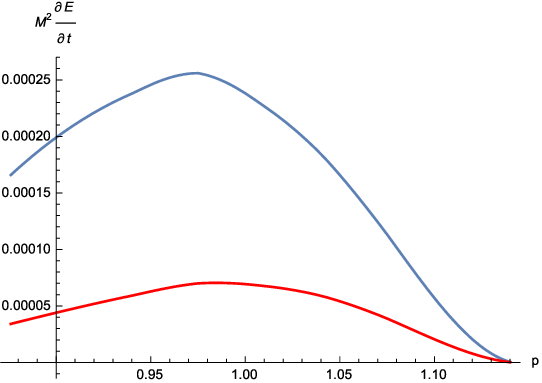}}
\caption{Energy emission rates for Dirac (top, blue) and Maxwell (bottom, red) fields.}\label{fig:5}
\end{figure}

However, since the mass of the black hole decreases as $p$ is increased, it is interesting to compare the intensities of Hawking radiation for Schwarzschild and non-Schwarzschild black holes \emph{of the same mass}. For spherically symmetric black holes the Hawking temperature is usually the dominant factor determining the intensity of Hawking radiation, because it is the argument of the exponent in the Hawking formula (\ref{energy-emission-rate}), while grey-body factors are only linear factors. From fig. (\ref{fig:Temperature}) one can see that the temperature monotonically decreases when $p$ is increased, approaching zero in the extremal limit. Then, one would expect that the non-Schwarzschild black hole will always be evaporating with a smaller rate than the Schwarzschild one. That is what we observe if we compare two black holes of the same radius (see fig. (\ref{fig:4a},\ref{fig:4b})), where one can also see that contribution of modes with $\ell > 3$ can be neglected. If we compare two black holes of the same mass, from fig. \ref{fig:5} one can see that at small and moderate $p$ the energy emission rate is increased and only at sufficiently large $p$ the growth of intensity is changed by the fall. This way, there is a regime of small and moderate $p$ for which the black hole is getting cooler and at the same time radiates at a higher rate. This happens because of the influence of grey-body factors and the decreased mass of the black hole: at not very large $p$ the Hawking temperature decays slowly (see fig. (\ref{fig:Temperature})), while the effective potential becomes much lower in the units of black-hole mass $M$ (see fig. (\ref{fig:1A},\ref{fig:1B})). Therefore, much larger transmission coefficient enhances Hawking radiation more than the decaying temperature suppresses it. At sufficiently small temperatures, the exponential factor of temperature dominates over the linear factor of the transmission coefficient and the emission becomes strongly suppressed.


\begin{table*}
\begin{tabular}{|c|c|c|c|c|c|c|}
  \hline
  \hline
  $p$ & $M$ & $dE/dt$ (Dirac) &  $dE/dt$ (Maxwell) &  $\tau_{1} (M_{0}/kg)^{3} $  &  $\tau_2 (M_{0}/kg)^{3}$ & Ultra-relativistic $M$ (in $kg.$)\\
   \hline
  0.876& 0.500252 & 0.00066 & 0.00014 & $8.7 \cdot 10^{-18}$ & $4.7 \cdot 10^{-18}$ & $5 \cdot 10^{11}  \ll M \ll 10^{13}   $\\
    \hline
  0.909& 0.441602 & 0.00107 & 0.00024 & $6.8 \cdot 10^{-18}$ & $3.8 \cdot 10^{-18}$ & $4 \cdot 10^{11}  \ll M  \ll  9 \cdot 10^{13}   $\\
   \hline
  0.942& 0.382094 & 0.00164 & 0.00041 & $5.8 \cdot 10^{-18}$  & $3.2 \cdot 10^{-18}$  & $3.8 \cdot 10^{11}  \ll M \ll 8 \cdot 10^{13}  $\\
  \hline
  0.975& 0.321730 & 0.00239 & 0.00065 & $5.5 \cdot 10^{-18}$ & $3 \cdot 10^{-18}$ &  $3 \cdot 10^{11}  \ll M \ll 7 \cdot 10^{13}   $ \\
   \hline
  1.008& 0.260509 & 0.00337 & 0.00100 & $5.9 \cdot 10^{-18}$  & $3.3 \cdot 10^{-18}$  &  $3 \cdot 10^{11}  \ll M \ll 6 \cdot 10^{13}  $ \\
    \hline
  1.041& 0.198432 & 0.00464 & 0.00149 & $7.2 \cdot 10^{-18}$ & $4.1 \cdot 10^{-18}$ &  $2 \cdot 10^{11} \ll M \ll 5 \cdot 10^{13}   $ \\
    \hline
  1.074& 0.135497 & 0.00629 & 0.00216 & $1.1 \cdot 10^{-17}$ & $6.4 \cdot 10^{-18}$ & $1.6 \cdot 10^{11}  \ll M \ll 3.4 \cdot 10^{13}   $\\
   \hline
  1.107& 0.071706 & 0.00834 & 0.00305 & $3 \cdot 10^{-17}$ & $1.7 \cdot 10^{-17}$ & $10^{11}  \ll M \ll 2 \cdot 10^{13}   $\\
   \hline
  1.140& 0.007058 & 0.01092 & 0.00425 & $7.6 \cdot 10^{-7}$ & $1.3 \cdot 10^{-6}$ & $10^{10}  \ll M \ll 2 \cdot 10^{12}   $\\
   \hline
  \hline
\end{tabular}
\caption{Energy emission rates for Maxwell and Dirac particles (in the units $r_{0}=1$), lifetimes of the black hole in the regime of negligible radiation of massive particles $\tau_{1}$ and in the ultra-relativistic regime $\tau_{2}$. In the last column the range of black-hole masses corresponding to the ultra-relativistic regime is written down.}\label{table2}
\end{table*}

According to Don Page \cite{Page:1976df} there are two different regimes of emission of particles: the first one, when the black hole mass is large enough and radiation of massive particles can be neglected. In this regime the radiation occurs mainly due to massless electron and muon neutrinos, photons, and gravitons. When the black-hole mass is sufficiently small, emission of electrons and positrons will occur ultra-relativistically. In the latter case, their law of radiation can be approximated by that for the massless Dirac field and the emission rate of all the Dirac particles should then be simply doubled. Supposing that the peak in the Dirac particles' spectrum  $\partial^{2} E/\partial t \partial \omega$ occurs at some $\omega \approx \xi M^{-1} $, we have the range of ultra-relativistic radiation of massive particles determined as follows:
\begin{equation}\nonumber
m_{e} = 4.19 \times 10^{-23} m_{p} \ll \xi M^{-1} \ll m_{\mu} = 8.65 \times 10^{-21} m_{p}.
\end{equation}
This inequality can be rewritten in the following form:
\begin{equation}
\xi^{-1} \cdot 10^{11} kg. \ll M \ll \xi^{-1} \cdot 2 \times 10^{12} kg.
\end{equation}
The maximum of the spectrum shifts towards smaller $\omega M$ when $p$ is increased. For example, we have for $p= 0.876$, $\xi = 0.18$; $p=0.975$,  $\xi = 0.14$; $p=1.041 $,  $\xi = 0.09$; $p=1.140  $,  $\xi = 0.004$.

The energy emitted causes the black-hole mass to decrease at the following rate \cite{Page:1976df}
\begin{equation}
\frac{d M}{d t} = -\frac{\hbar c^4}{G^2} \frac{\alpha_{0}}{M^2},
\end{equation}
where we have restored the dimensional constants. Here $\alpha_{0} = d E/d t$ is taken for a given initial mass $M_{0}$. Since most of its time the black hole spends near  its original state $M_{0}$ and integrating of the above equation  gives us the life-time of a black hole:
\begin{equation}
\tau = \frac{G^2}{\hbar c^4} \frac{M_{0}^3}{3 \alpha_{0}}.
\end{equation}

From the table (\ref{Table1}) one can see that the contribution of gravitons in the total energy emission is less than two percents, so that for the robust evaluation of the black-hole lifetime it can be neglected. Strictly speaking, the same is not guaranteed for the non-Schwazrschild black hole which we consider. However, since the reduction of gravitational perturbations to the master-like form in the general case is a complicated problem we will not consider emission of gravitons here.  This should provide a robust estimation of the evaporation process at least when the coupling constant is far from its extremal value. From the table (\ref{table2}) we can see that the lifetime of black holes with mass which is much larger than $10^{12}-10^{13} kg.$ is increased by 11 orders at $p=1.14$. At the same time, the range of masses of black holes within which the ultra-relativistic emission cannot be neglected does not change much, shifting relatively moderately towards smaller masses.

\section{Conclusion}

Semiclassical regime of Hawking radiation in the Einstein theory of gravity was studied in a great number of papers (see, for example, \cite{Page:1976df,Konoplya:2014sna,Kokkotas:2010zd,Boonserm:2014rma,Pappas:2017kam} and references therein). To the best of our knowledge Hawking radiation in the four-dimensional alternative theories of gravity, especially those with higher curvature corrections were, mostly, not studied so far. In the present paper we have considered Hawking radiation in the vicinity of non-Schwarzschild black hole in the higher derivative theory of gravity, which includes the quadratic (Weyl) correction to the Einstein action. We have found that the evaporation of non-Schwarzschild black holes is essentially different from the Schwarzschild ones:
\begin{itemize}
\item Although the temperature of the black hole monotonically decreases when the coupling $p$ is increased, the Hawking evaporation is enhanced up until moderate values of $p$. At larger $p$ the growth of the Hawking evaporation rate is changed by the fall off. This feature was unexpected and can be explained by the role of grey-body factors in the evaporation process: at relatively small $p$, when the temperature decreases still slowly, the height of the effective potential for the black hole of the same mass is much lower than its Schwarzschild analogue, allowing, thereby, for much higher transmission coefficient for radiation.

\item In the extremal limit when the mass of the black hole goes to zero, the energy emission rate vanishes.

\item The range of black-hole masses for which massive electrons and positrons are emitted in the ultra-relativistic regime at a considerable rate is shifted towards smaller masses for sufficiently large values of $p$.
\end{itemize}

The question which was beyond the scope of our work is the role of gravitons in the black hole evaporation. For the Schwazrschild black hole it is evident that the contribution of gravitons does not exceeds 2 $\%$ of the total energy emission. However, this is not guaranteed for the non-Schwarzschild solution and we hope that future publications will check whether contribution of gravitons could be neglected in this case as well.

\acknowledgments{
The authors acknowledge  the  support  of  the  grant  19-03950S of Czech Science Foundation ($GA\check{C}R$). This publication has been prepared with the support of the “RUDN University Program 5-100”. AZ acknowledges the SU grant SGS/12/2019.
}

\newpage
\appendix

\begin{widetext}
\section{Analytical form of the metric functions}\label{sec:metricform}

The metric coefficients are determined as follows:

\begin{equation}
e^{\mu} = \left(1-\frac{r_0}{r}\right)A(r), \qquad e^{\nu} = \frac{B(r)^2}{\left(1-\frac{r_0}{r}\right)A(r)},
\end{equation}
where

\begin{eqnarray}\nonumber
A(r)&=&\Biggr[152124199161 \left(873828 p^4-199143783 p^3+806771764 p^2-1202612078 p+604749333\right) r^4
\\\nonumber&&+78279\left(1336094371764p^6-300842119184823 p^5+393815823540843 p^4+2680050514097926 p^3\right.
\\\nonumber&&\left.-9501392159249689 p^2+10978748485369369 p-4249747766121792\right)r^3 r_0
\\\nonumber&&-70372821 \left(1486200636 p^6+180905642811 p^5+417682197141 p^4-1208134566031 p^3\right.
\\\nonumber&&\left.-324990706209 p^2+3382539200269p-2557857695019\right) r^2 r_0^2-\left(104588131327314156 p^6\right.
\\\nonumber&&-23549620247668759617 p^5-435688050031083222417p^4+2389090517292988952355 p^3
\\\nonumber&&\left.-3731827099716921879958 p^2+2186684376605688462974 p-389142952738481370396\right) r r_0^3
\\\nonumber&&+31\left(3373810687977876 p^6+410672271594465801 p^5-14105000476530678231 p^4+51431640078486304191 p^3\right.
\\
&&\left.-71532183052581307042p^2+43250367615320791700 p-9476049523901501640\right) r_0^4\Biggr]
\\\nonumber&&/\Biggr[152124199161 r^2\Biggr(\left(873828 p^4-199143783 p^3+806771764 p^2-1202612078 p+604749333\right) r^2
\\\nonumber&&-2 \left(873828 p^4-47583171 p^3+386036980 p^2-678598463 p+341153481\right) r r_0
\\\nonumber&&+899\left(972 p^4+115659 p^3-38596 p^2-1127284 p+1101579\right) r_0^2\Biggr)\Biggr]\,,
\end{eqnarray}
\begin{eqnarray}\nonumber
B(r)&=&\Biggr[464405 \left(3251230164 p^3-14548777134 p^2+20865434326 p+23094914865\right) r^3
\\\nonumber&&-464405 \left(6502460328 p^3-52856543928p^2+100077612184 p-32132674695\right) r^2 r_0
\\\nonumber&&-\left(1244571650887908 p^3+17950319416564777 p^2-53210739821255918p+5097428297648940\right) r r_0^2
\\
&&+635371 \left(4335198168 p^3-42352710803 p^2+90235778452 p-49464019740\right)r_0^3\Biggr]
\\\nonumber&&/\Biggr[464405 r \Biggr(\left(3251230164 p^3-14548777134 p^2+20865434326 p+23094914865\right) r^2
\\\nonumber&&-\left(6502460328p^3-52856543928 p^2+100077612184 p-32132674695\right) r r_0
\\\nonumber&&+6\left(541871694 p^3-6384627799 p^2+13202029643p+2626009760\right) r_0^2\Biggr)\Biggr]\,.
\end{eqnarray}
\end{widetext}


\end{document}